\documentclass
[preprint,superbib,superscriptaddress,byrevtex,showpacs,prb]{revtex4}%
\usepackage{amsfonts}
\usepackage{amsmath}
\usepackage{amssymb}
\usepackage{graphicx}%
\begin{document}
\preprint{ }
\title[Short title for running header]{$^{31}$P NMR study of Na$_{2}$CuP$_{2}$O$_{7}$: a $S=1/2$ two-dimensional
Heisenberg antiferromagnetic system }
\author{R. Nath}
\affiliation{Department of Physics, Indian Institute of Technology, Mumbai 400076, India.}
\author{A. V. Mahajan}
\affiliation{Department of Physics, Indian Institute of Technology, Mumbai 400076, India.}
\author{N. B\"{u}ttgen}
\affiliation{Experimentalphysik V, Elektronische Korrelationen und Magnetismus, Institut
f\"{u}r Physik, Universit\"{a}t Augsburg, D-86135 Augsburg, Germany. }
\author{C. Kegler}
\affiliation{Experimentalphysik V, Elektronische Korrelationen und Magnetismus, Institut
f\"{u}r Physik, Universit\"{a}t Augsburg, D-86135 Augsburg, Germany. }
\author{J. Hemberger}
\affiliation{Experimentalphysik V, Elektronische Korrelationen und Magnetismus, Institut
f\"{u}r Physik, Universit\"{a}t Augsburg, D-86135 Augsburg, Germany. }
\author{A. Loidl}
\affiliation{Experimentalphysik V, Elektronische Korrelationen und Magnetismus, Institut
f\"{u}r Physik, Universit\"{a}t Augsburg, D-86135 Augsburg, Germany. }

\begin{abstract}
The magnetic properties of Na$_{2}$CuP$_{2}$O$_{7}$ were investigated by means
of $^{31}$P nuclear magnetic resonance (NMR), magnetic susceptibility, and
heat capacity measurements. We report the $^{31}$P NMR shift, the spin-lattice
$\left(  1/T_{1}\right)  $, and spin-spin $\left(  1/T_{2}\right)  $
relaxation-rate data as a function of temperature $T$.\ The temperature
dependence of the NMR shift $K(T)$ is well described by the $S=1/2$ square
lattice Heisenberg antiferromagnetic (HAF) model with an intraplanar exchange
of $J/k_{B}$ $\simeq$ ($18\pm2$) K and a hyperfine coupling $A$ =
($3533\pm185$) Oe/$\mu_{B}$. The $^{31}$P NMR spectrum was found to broaden
abruptly below $T\sim$ $10$ K signifying some kind of transition. However, no
anomaly was noticed in the bulk susceptibility data down to $1.8$ K. The heat
capacity appears to have a weak maximum around $10$ K. With decrease in
temperatures, the spin-lattice relaxation rate $1/T_{1}$ decreases
monotonically and appears to agree well with the high temperature series
expansion expression for a $S$ =$1/2$ 2D square lattice.

\end{abstract}
\volumeyear{year}
\volumenumber{number}
\issuenumber{number}
\eid{identifier}
\date[Date text]{date}
\received[Received text]{date}

\revised[Revised text]{date}

\accepted[Accepted text]{date}

\published[Published text]{date}

\startpage{1}
\endpage{ }
\maketitle

\section{\textbf{Introduction}}

Low-dimensional spin systems with antiferromagnetic interactions have received
considerable attention because of pronounced quantum mechanical effects which
result in magnetic properties which are quite different from those of the
three-dimensional (3D) antiferromagnetic substances. According to the
Hohenberg-Mermin-Wagner theorem\cite{mermin1966}, one-dimensional (1D) and
two-dimensional (2D) spin systems with Heisenberg interaction and finite-range
coupling between spins cannot have long-range order (LRO) at any temperature
different from zero. In 1D systems it does not occur even at zero temperature.
However for 2D systems at zero temperature, LRO is not forbidden by this
theorem. In fact, LRO has been rigorously established for spin $S>1/2$, on a
square lattice with nearest-neighbor coupling.\cite{affleck1988,neves1986} For
the case of $S=1/2$, there is no solid proof but there are strong theoretical
arguments that LRO exists.\cite{kennedy1988,singh1988}

The case of a $S$ = $1/2$ system on a square lattice, with nearest-neighbor
antiferromagnetic coupling, has been of special interest because of its
proximity to the high-$T_{c}$ cuprates. The interest in quasi-1D spin systems
has been stimulated by the hope that better understanding of 1D systems might
lead to insights into the 2D systems and high-temperature superconductivity.

So far, a large number of $S=1/2$ compounds have been experimentally
investigated, which could effectively be described by 2D HAF models. Among the
important ones are cuprate compounds such as La$_{2}$CuO$_{4}$ and YBa$_{2}%
$Cu$_{3}$O$_{6}$ which have CuO planes.\cite{aeppli1989,singh1989a,wu1987} In
La$_{2}$CuO$_{4}$, the intraplanar exchange coupling $J/k_{B}$ was reported to
be $1800$ K. In spite of a small interplanar coupling $J%
\acute{}%
\approx10^{-5}J$, the large correlations in CuO planes lead to 3D magnetic
order at the N\'{e}el temperature $T_{N}$ $\sim$ $300$ K. In the above cases
$T_{N}$ is strongly dependent on $J%
\acute{}%
/J$ ratio.

Most recently, KCuF$_{3}$ has been experimentally investigated as one of the
quasi-1D HAF compounds.\cite{chakhalian2003,hutchings1969} Unfortunately, this
material has a relatively large coupling ratio $J%
\acute{}%
/J\sim1.0\times10^{-2}$, due to which, the $T_{N}/J$ ratio ($\sim39$ K/$203$
K) was found to be large. Later Sr$_{2}$CuO$_{3}$ was reported to be another
1D HAF system,\cite{motoyama1996} with a significantly reduced $T_{N}/J$ ratio
of $\sim5$ K/$2200$ K $\sim2\times10^{-3}$. Due to the relatively small
$T_{N}$ value compared to the exchange coupling $J/k_{B}$, 1D behavior is
observed over a wide range of temperature. Recently, Nath \textit{et
al}.\cite{nath2004} have reported that Sr$_{2}$Cu(PO$_{4}$)$_{2}$ and Ba$_{2}%
$Cu(PO$_{4}$)$_{2}$ are two 1D HAF systems, which do not appear to undergo
N\'{e}el ordering even at a very low temperature ($T\sim0.02$ K). Their
exchange coupling constants have been reported to be $165$ K, and $151$ K
respectively. It is thus of interest to synthesize and characterize additional
$S=1/2$ 1D or 2D HAF compounds to improve our understanding of such systems.
In this paper, we present a detailed study of the magnetic properties of
Na$_{2}$CuP$_{2}$O$_{7}$ via susceptibility and $^{31}$P nuclear magnetic
resonance (NMR) experiments. Our objective is to report detailed magnetic
measurements on a new potentially low-dimensional system and to analyse the
data based on available models. \ An additional objective is then to motivate
the theorists to model real (and more complex) systems as the ones we report
and compare the result of their simulations with our data. \ Our results
indicate that the magnetic properties of Na$_{2}$CuP$_{2}$O$_{7}$ are
consistent with the 2D square-lattice S = 1/2 HAF model somewhat better than
the 1D model. \ In the next section, an overview of the schematic structure is
presented which motivated us to work on this system. \ This is followed by
details of our experiments. \ The "experimental details" section is followed
by results of our magnetic and heat capacity measurements accompanied by an
analysis and a conclusion.

\section{Structure}

The structural properties of Na$_{2}$CuP$_{2}$O$_{7}$ have been reported by
Etheredge \textit{et al.}\cite{kristen1995} and Erragh \textit{et
al.}\cite{fatima1995} It (the high-temperature phase) crystallizes in a
monoclinic unit cell with space group C$_{2/c}$. The reported lattice
constants are $14.715$ \AA , $5.704$ \AA , and $8.066$ \AA respectively along
$a$-, $b$- and $c$- directions. From the schematic diagram of the structure
shown in Fig. 1 (a) and (b), it is seen that the exchange interaction between
Cu$^{2+}$ ions could arise due to two interaction paths: (i) each CuO$_{6}$
octahedron shares its corners with two similar kinds of PO$_{4}$ groups. The
corner sharing takes place in one direction forming [Cu(PO$_{4}$)$_{2}%
$]$_{\infty}$ chains along the $c$-direction and in this case the magnetic
properties would be those of a 1D HAF chain. (ii) alternatively, there exist
(nearly) $180^{\circ}$ Cu-O-Cu linkages in the $bc$ plane. Depending on the
orientation of the $d_{x^{2}-y^{2}}$ orbitals (\textit{i.e.} whether
perpendicular to the $bc$ plane or in the $bc$ plane), the system will behave
either as a chain-like or a planar magnetic system. In the latter case
however, one should note that the Cu$^{2+}$ ions are arranged in a face
centered manner (see Fig. 1(b)). One can, therefore, think of the $bc$ plane
as comprising of two sub-planes (shown in Fig. 1(b) by thick and thin lines).
While the Cu$_{1}-$O$-$Cu$_{1}$ (or Cu$_{2}-$O$-$Cu$_{2}$) bond angle is
almost $180%
{{}^\circ}%
$, the Cu$_{1}-$O$-$Cu$_{2}$ bond angle is somewhat less than $90%
{{}^\circ}%
$. The deviation of the magnetic properties of Na$_{2}$CuP$_{2}$O$_{7}$ from
the square planar case would depend on the relative strength of the Cu$_{1}%
-$O$-$Cu$_{2}$ interaction with respect to that of the Cu$_{1}-$O$-$Cu$_{1}$
(or Cu$_{2}-$O$-$Cu$_{2}$) interaction. Further, the interaction between the
$bc$ planes is expected to be weak since the distance between them is about
$8$ \AA which is nearly twice the Cu$^{2+}-$Cu$^{2+}$ intraplanar distance of
$4$ \AA . Also, unlike the $180%
{{}^\circ}%
$ Cu$_{1}-$O$-$Cu$_{1}$ intraplanar bonds, there appear to be no similar
interaction paths perpendicular to the $bc$ plane.

\section{Experimental details}

Polycrystalline sample of Na$_{2}$CuP$_{2}$O$_{7}$ was prepared by solid state
reaction technique using NaH$_{2}$PO$_{4}$.H$_{2}$O (98\% pure) and CuO
(99.99\% pure) as starting materials. The stoichiometric mixtures were fired
at 800 $^{\circ}$C for 120 hours, in air, with several intermediate grindings
and pelletization. Formation of nearly single phase sample was confirmed from
x-ray diffraction, which was performed with a Philips\ Xpert-Pro powder
diffractometer. A Cu target was used in the diffractometer with $\lambda
_{av}=1.54182$ \AA . An impurity phase was identified to be Na$_{3}$P$_{3}%
$O$_{9}$ and the intensity ratio ($I_{imp}/I_{max}$) was found to be 0.05,
where $I_{imp}$ is the intensity of the most intense diffraction peak for
Na$_{3}$P$_{3}$O$_{9}$ and $I_{max}$ is that of Na$_{2}$CuP$_{2}$O$_{7}$.
Lattice parameters were calculated using a least-square fit procedure. The
obtained lattice constants are $14.703$ (4) \AA , $5.699$ (2) \AA , and
$8.061$ (3) \AA , respectively along $a$-, $b$- and $c$- directions. These are
in agreement with previously reported values. Magnetization ($M$) data were
measured as a function of temperature $T$ ($1.8$ K $\leq$ $T$ $\leq$ $400$ K)
and applied field $H$ ($0$ $\leq$ $H$ $\leq$ $50$ kG) using a SQUID
magnetometer (Quantum Design). The heat capacity was measured with a PPMS
set-up (Quantum Design). The NMR measurements were carried out using pulsed
NMR techniques on $^{31}$P nuclei (nuclear spin $I=1/2$ and gyromagnetic ratio
$\gamma/2\pi=17.237$ MHz/Tesla) in a temperature range $2$ K $\leq$ $T$ $\leq$
$300$ K using a $^{4}$He cryostat (Oxford Instruments). We have done the
measurements in an applied field of about $55$ kG, which corresponds to a
radio frequency (rf) of about $95$ MHz. Spectra were obtained by plotting the
echo integral (following a $\pi/2-\pi$ pulse sequence with a $\pi/2$ pulse of
width $4$ $\mu s$) as a function of the field at a constant frequency of $95$
MHz. The NMR shift $K(T)=\left[  H_{ref}-H\left(  T\right)  \right]  /H(T)$
was determined by measuring the resonance field of the sample ($H\left(
T\right)  $) with respect to a reference H$_{3}$PO$_{4}$ solution (resonance
field $H_{ref}$). The $^{31}$P nuclear spin-lattice relaxation rate ($1/T_{1}%
$) was determined by the inversion-recovery method. Nuclear spin-spin
relaxation rate $\left(  1/T_{2}\right)  $ was obtained by measuring the decay
of the transverse nuclear magnetization with a variable spacing between the
$\pi/2$ and the $\pi$ pulse.

\section{Results and discussion}

We first present the results of our $^{31}$P NMR measurements in Na$_{2}%
$CuP$_{2}$O$_{7}$. Since there is a unique $^{31}$P site, the $^{31}$P NMR
spectra consist of a single spectral line at high temperatures ($T\geq5$ K) as
is expected for $I=1/2$ nuclei. The observed peak position shifts with respect
to $H_{ref}$ in the field-sweep spectra. The temperature dependence of the
$^{31}$P NMR shift apparently arises due to the temperature dependence of the
spin susceptibility $\chi_{spin}(T)$ via a hyperfine coupling to the Cu$^{2+}$
ions. The NMR shift is not affected by small amounts of extrinsic paramagnetic
impurities whereas in the bulk susceptibility they give rise to Curie terms.
NMR shift data as a function of temperature are shown in Fig. 2, for $5$ K
$\leq$ $T$ $\leq$ $300$ K. They exhibit a broad maximum at $20$ K, indicative
of short-range ordering. As explained earlier, the dominant magnetic behavior
of Na$_{2}$CuP$_{2}$O$_{7}$ could be that of a HAF chain or a plane.
Consequently, we tested both the 2D (planar) and 1D (chain) models to fit the
NMR shift data. A high temperature $(\frac{k_{B}T}{J}\gtrsim0.7)$ series
expansion for the inverse susceptibility $1/\chi_{spin}(T)$ for 2D $S$ = $1/2$
HAF square lattice was given by Rushbrooke and Wood\cite{rushbrooke1958},
which has the form,\textit{%
\begin{equation}
1/\chi_{spin}(T)=\frac{J}{N_{A}\mu_{B}^{2}g^{2}}\left[  4x+\sum_{n=1}^{6}%
\frac{C_{n}}{(\frac{4}{3}x)^{n-1}}\right] \label{2dHT}%
\end{equation}
}where $x=\frac{k_{B}T}{J}$, $g$ is the Land\'{e} $g$-factor, $\mu_{B}$ is the
Bohr magneton, $N_{A}$ is the Avogadro number, and $C_{n}$ are the
coefficients listed in Table 1 of Ref. 13. Similarly
Johnston\cite{johnston1996} parametrized the low-temperature $\left(
\frac{k_{B}T}{J}\leq1\right)  $ simulations of Takahashi\cite{takahashi1989}
and Makivic and Ding\cite{makivic1991} to obtain \textit{%
\begin{equation}
\chi_{spin}(T)=\frac{N_{A}\mu_{B}^{2}g^{2}}{J}\left[
0.043669+0.039566x-0.5341x^{3}+4.684x^{4}-11.13x^{5}+10.55x^{6}-3.56x^{7}%
\right] \label{2dLT}%
\end{equation}
}For 1D HAF chains, the temperature dependence of the susceptibility
$\chi_{spin}\left(  T\right)  $ was numerically\ calculated by Bonner and
Fisher \cite{bonner1964} which accurately predicts the susceptibility for high
temperatures ($\frac{k_{B}T}{J}\geq0.5$). \ Below, we use the form as given by
Estes \textit{et al.} \cite{estes1978}. \textit{%
\begin{equation}
\chi_{spin}(T)=\frac{Ng^{2}\mu_{B}^{2}}{k_{B}x}\times\frac{\left(
0.25+0.074975x^{-1}+0.075235x^{-2}\right)  }{\left(  1+0.9931x^{-1}%
+0.172135x^{-2}+0.757825x^{-3}\right)  }\label{BF}%
\end{equation}
}

Since the temperature dependence of the $\chi_{spin}(T)$ is reflected in the
NMR shift $K(T)$, one can determine the exchange coupling $J/k_{B}$ and the
hyperfine interaction $A$ simultaneously by fitting the temperature dependence
of $K$ to the following equation,

\textit{%
\begin{equation}
K(T)=K_{0}+\left(  \frac{A}{N_{A}\mu_{B}}\right)  \chi_{spin}\left(  T\right)
\label{shift}%
\end{equation}
}where $K_{0}$ is the chemical shift. Figs. 2 and 3 show fitting of the
$^{31}$P NMR shift data to Eq. \ref{shift} taking $\chi_{spin}$ for HAF square
lattice (Eq. \ref{2dHT}) and linear chain (Eq. \ref{BF}) respectively. In Fig.
2, the fitting to the 2D high-temperature series expansion was done for $15$ K
$\leq T\leq300$ K whereas in Fig. 3, the experimental data were fitted to the
linear chain model in the temperature range $5$ K $\leq T\leq300$ K. The
parameters extracted from the fit are listed in Table I. The Land\'{e}
\textit{g}-value was found to be $g=2.1$, which is a typical value for cuprates.%

\begin{table}[h] \centering
\caption{Values of the parameters obtained from the fitting of NMR shift to Eq. 4 considering 2D square planar HAF (Eq. 1) and 1D HAF (Eq. 3) models.}%
\begin{tabular}
[c]{cccc}\hline\hline
& $\hspace{0.5cm}K_{0}$ (ppm)\hspace{0.5cm} & $\hspace{0.5cm}A$ (Oe/$\mu_{B}%
$)\hspace{0.5cm} & $\hspace{0.5cm}\frac{J}{k_{B}}$ (K)\hspace{0.5cm}\\\hline
Fig. 2 & $100\pm50$ & $3533\pm185$ & $18\pm2$\\
Fig. 3 & $143\pm60$ & $3479\pm200$ & $28\pm5$\\\hline\hline
\end{tabular}%
\end{table}%

From Figs. 2 and 3 it is seen that our $K(T)$ data fit somewhat better to the
2D HAF model. At low temperature ($T\leq25$ K), the 1D fit deviates from the
experimental data. In Fig. 2, we have also plotted the simulated low
temperature curve using Eqs. \ref{2dLT} and \ref{shift} with the parameters
$K_{0}$, $A,$ and $J/k_{B}$ obtained from the high temperature fit along with
our experimental data. It is clearly seen that our experimental data do not
deviate significantly from the simulated curve down to $T\sim10$ K while a
large deviation is seen below $10$ K. This suggests some transition or
crossover below $10$ K.

An observation of the $^{31}$P NMR lineshapes below $10$ K (shown in Fig. 4)
reveals a huge broadening at lower temperatures. Further, the lineshape
develops shoulder-like features and finally at $2$ K the overall extent of the
spectrum is about five times that at 10 K, with at least three distinct peaks.
Either a structural or a magnetic transition might be the cause for this. No
anomaly is seen in the bulk susceptibility (see below) which seems to suggest
against the occurrence of 3D LRO.

Magnetic susceptibility $\chi(T)$ (= $M/H$) of Na$_{2}$CuP$_{2}$O$_{7}$ was
measured as a function of temperature in an applied field of $5$ kG (Fig. 5).
The amount of ferromagnetic impurity present in our sample was estimated from
the intercept of $M$ vs. $H$ isotherms at various temperatures and was found
to be $19$ ppm of ferromagnetic Fe$^{3+}$ ions. The data in Fig. 5 have been
corrected for these ferromagnetic impurities. As shown in the figure,
$\chi(T)$ exhibit a broad maximum at $20$ K, indicative of low-dimensional
magnetic interactions. With a further decrease in temperature, susceptibility
increases in a Curie-Weiss manner. This possibly comes from defects and
extrinsic paramagnetic impurities present in the samples. No obvious features
associated with LRO are seen for $1.8$ K $\leq$ $T$ $\leq$ $400$ K.

The broad maximum in $\chi(T)$ at $20$ K could be reproduced by assuming that

\textit{%
\begin{equation}
\chi=\chi_{0}+\frac{C}{T+\theta}+\chi_{spin}(T)\label{chi}%
\end{equation}
}where $\chi_{spin}(T)$ is the uniform spin susceptibility for a $S=1/2$ 2D
HAF system obtained from Eq. \ref{2dHT}. $\chi_{0}$ is temperature independent
and consists of diamagnetism of the core electron shells ($\chi_{core}$) and
Van-Vleck paramagnetism($\chi_{vv}$) of the open shells of the Cu$^{2+}$ ions
present in the sample. The Curie-Weiss contribution is $\frac{C}{T+\theta}$
$\left(  \text{where }C=\frac{Ng^{2}\mu_{B}^{2}S(S+1)}{3k_{B}}\right)  $due to
paramagnetic species in the sample. The parameters were determined by fitting
our experimental $\chi(T)$\ data to Eq. \ref{shift} in the high temperature
regime $15$ K $\leq$ $T$ $\leq$ $400$ K. The extracted parameters are
$\chi_{0}=$($-7\pm2$) $\times10^{-5}$ cm$^{3}$/mole, $C=$($13\pm3$%
)$\times10^{-3}$ cm$^{3}$K/mole, $\theta=1.7$ K, $\frac{J}{k_{B}}=$ ($18\pm2$)
K, and $g=2.07$. Adding the core diamagnetic susceptibility for the individual
ions\cite{Magnetochemistry} (Na$^{1+}$ = -5 $\times$ 10$^{-6}$ cm$^{3}$/mole,
Cu$^{2+}$ = -11 $\times$ 10$^{-6}$ cm$^{3}$/mole, P$^{5+}$ = -1 $\times$
10$^{-6}$ cm$^{3}$/mole, O$^{2-}$ = -12 $\times$ 10$^{-6}$ cm$^{3}$/mole), the
total $\chi_{core}$ was calculated to be $-1.07\times10^{-4}$ cm$^{3}$/mole.
The Van-Vleck paramagnetic susceptibility estimated by subtracting
$\chi_{core}$ from $\chi_{0}$ is about $3.7\times10^{-5}$ cm$^{3}$/mole, which
is comparable to that found in Sr$_{2}$CuO$_{3}$ ($\sim$ 3.4$\times$10$^{-5}$
cm$^{3}$/mole)\cite{motoyama1996}. The Curie contributions present in the
sample corresponds to a defect spin concentration of $3.5$ \% assuming defect
spin $S=1/2$.

Further, we did heat capacity measurements on Na$_{2}$CuP$_{2}$O$_{7}$ to look
for signs of any anomalies at low temperature, signaling a magnetic
transition. \ As seen in the data (Fig. 6), \ no sharp peaks are visible which
seems to rule out LRO. \ However, a look at the derivative of the specific
heat as a function of temperature (see inset of Fig. 6) clearly shows a local
maximum at about $7$ K and a local minimum around $10$ K. \ This suggests that
the specific heat has an anomaly/peak between $7$ K and $10$ K.\cite{footnote}%
\ In the 2D HAF model, a broad maximum in the heat capacity is expected at
about $T=0.58J/k_{B}$ \cite{makivic1991} (i.e. at about $10$ K in the present
case) whereas in the 1D HAF\ model, a broad maximum is expected at about
$T=0.48J/k_{B}$\cite{johnston2000} (i.e. at about $16$ K in the present case).
\ This further suggests the applicability of the 2D HAF model in the present case.

The time dependence of the longitudinal nuclear magnetization $M(t)$ for
$^{31}$P at three different temperatures is shown in the inset of Fig. 7. For
a spin-1/2 nucleus this recovery is expected to follow a single exponential
behavior,
\begin{equation}
\frac{M\left(  \infty\right)  -M\left(  t\right)  }{M\left(  \infty\right)
}=A\exp\left(  -\frac{t}{T_{1}}\right)  +C \label{T1}%
\end{equation}
Our experimental data show good single exponential behaviour over two decades.
The spin-lattice relaxation rate $1/T_{1}$ was extracted from the fitting of
the experimental data at various temperatures (down to $5$ K) to Eq. \ref{T1}
. Due to the large line broadening we could not saturate the nuclear
magnetization below $5$ K and hence could not extract reliable $1/T_{1}$ below
this temperature. The temperature dependence of $^{31}$P nuclear spin-lattice
relaxation rate thus obtained is presented in Fig. 7. With decrease in
temperature, it decreases monotonically. For $S=1/2$ 2D square lattice, a high
temperature series expansion for $1/T_{1}$ was given by
Moriya\cite{moriya1956} which has the form,
\begin{equation}
1/T_{1}=\frac{1}{T_{1\infty}\left(  1+J/(4k_{B}T)\right)  ^{1/2}\exp\left[
(J/(2k_{B}T))^{2}(1+J/(4k_{B}T))\right]  } \label{2Dt1}%
\end{equation}
where, $\frac{1}{T_{1\infty}}=(\frac{A_{th}^{2}}{\hbar^{2}})(\frac{\sqrt{\pi
}\hbar k_{B}}{4J})$ taken from Ref. 20. Here, $A_{th}=2A\hbar\gamma$ where $A$
is the total hyperfine coupling obtained from the experiment. In Moriya's
expression a term of the order of $(J/k_{B}T)^{2}$ occurs in the prefactor and
has a negligible effect and hence is not included in Eq. \ref{2Dt1}. Using Eq.
\ref{2Dt1} and the relevant $A$ and $J/k_{B}$ values obtained from the fit of
NMR shift to 2D model, we simulated the theoretical curve for Na$_{2}$%
CuP$_{2}$O$_{7}$ and it is plotted with our experimental data in Fig. 7.
Clearly, at high temperatures ($T\geq25$ K), the experimental data agree
resonably well with the simulated curve. \ In the case of a $S=1/2$ 1D HAF
chain model, $1/T_{1}$ is expected to be temperature independent at low
temperatures ($T<<J/k_{B}$) due to a dominant contribution from the
fluctuations of the staggered susceptibility of the 1D chain. \ We are unable
to probe this region of temperature due to the large broadening of our NMR
spectra seen there. \ At higher temperatures, from fluctuations of the uniform
susceptibility of the 1D chain, one expects a linear variation of $1/T_{1}%
$with temperature. This should eventually saturate at even higher temperatures
when the spin-susceptibility becomes Curie-like. \ Qualitatively speaking, the
observed $1/T_{1}$ data could be explained based on the above. \ However, an
analytical expression for the temperature dependence of $1/T_{1}$ is not
available in the temperature regime of our experiment and hence no curve-fit
is shown in the figure.

The spin-spin relaxation was measured as a function of separation time $t$
between $\pi/2$ and $\pi$ pulses by monitoring the decay of the transverse
magnetization. $1/T_{2}$ at different temperatures was obtained by fitting of
the spin-echo decay to the following equation,%

\begin{equation}
M\left(  2t\right)  =M_{0}\exp\left[  -2\left(  \frac{t}{T_{2}}\right)
\right]  +C \label{T2}%
\end{equation}
Inset of Fig. 8 shows the spin-echo decays for different temperatures. The
extracted spin-spin relaxation rates $1/T_{2}$ are plotted as a function of
temperature in Fig. 7. It is to be seen that below about $55$ K, the spin-spin
relaxation rate $1/T_{2}$ falls sharply towards low temperatures. No
indication of 3D LRO was found down to 6 K. The origin of this temperature
dependence $1/T_{2}$ is not clear yet.

\section{\textbf{Conclusion}}

Our $^{31}$P NMR shift and susceptibility data fitted reasonably well to the
high temperature series expansion for 2D HAF model whereas the fitting to the
1D model was not as good, especially in the $T\leq25$ K regime. From the
$^{31}$P NMR shift analysis, the $J/k_{B}$ value was estimated to be about
($18\pm2$) K. The large broadening of the $^{31}$P NMR spectra at $T\leq5$ K
points towards a transition. However, no evidence of magnetic LRO was found in
susceptibility and heat capacity measurements down to 2 K. Further experiments
are required to really understand the detailed nature of this transition.
$^{31}$P NMR $1/T_{1}$ shows a good agreement with the theory of 2D HAF square
lattice. The results reported in this paper thus suggest a variety of
experiments and a need for a better theoretical understanding of quasi-low
dimensional Heisenberg antiferromagnets.

\begin{acknowledgments}
We would like to kindly acknowledge D. Vieweg for SQUID measurements and Th.
Wiedmann for heat capacity measurements. One of us (AVM) would like to thank
the Alexander von Humboldt foundation for financial support for the stay at
Augsburg. This work was supported by the BMBF via VDI/EKM, FKZ 13N6917-A and
by the Deutsche Forschungsgemeinschaft (DFG) through the
Sonderforschungsbereich SFB 484 (Augsburg).
\end{acknowledgments}

\textbf{Figure Captions}

\ FIG. 1 (a) \ A schematic figure of the $bc$ plane in Na$_{2}$CuP$_{2}$%
O$_{7}$ with $\left[  \text{Cu(PO}_{4}\text{)}_{\text{2}}\right]  _{\infty}$
linear chains propagating along $c$-direction indicated. A possible coupling
path Cu-O-P-O-Cu is also indicated. (b) The arrangement of Cu and O in the
$bc$-plane is shown. Two planes formed by Cu$_{1}^{2+}$ and Cu$_{2}^{2+}$ ions
are represented by thick and thin bonds respectively.

\ FIG. 2 \ $^{31}$P NMR shift $K$ vs temperature $T$ for Na$_{2}$CuP$_{2}%
$O$_{7}$. The solid line is fit to Eq. \ref{shift} in the temperature range,
$15$ K $\leq T\leq300$ K, where $\chi_{spin}$ is the high temperature series
expansion for susceptibility of HAF square plane (Eq. \ref{2dHT}). The dashed
line is the simulated curve from Eq. \ref{2dLT} using the parameters obtained
from the high-$T$ fit. In the inset we have displayed the data at low
temperatures on a logarithmic scale in order to show the deviation of
experimental data from low-$T$ series expansion.

\ FIG. 3 \ $^{31}$P NMR shift $K$ vs temperature $T$ for Na$_{2}$CuP$_{2}%
$O$_{7}$. The solid line is fit to Eq. \ref{shift} in the temperature range,
$5$ K $\leq T\leq300$ K taking $\chi_{spin}$ for HAF chain (Eq. \ref{BF}). In
the inset we have displayed the data on a logarithmic scale in order to show
the deviation of experimental data from the theory around the broad maximum region.

\ FIG. 4 \ Low-$T$ field sweep $^{31}$P NMR spectra for Na$_{2}$CuP$_{2}%
$O$_{7}$ are shown at different temperatures $T$ around 5 K. It also shows the
sudden change in line width and appearance of several distinct peaks.

\ FIG. 5 \ Magnetic susceptibility ($M/H$) vs temperature $T$ \ measured at
$H=5$ kG for Na$_{2}$CuP$_{2}$O$_{7}$. The solid line is best fit of the data
to Eq. \ref{chi} in the $15$ K $\leq T\leq400$ K range taking $\chi_{spin}$
for HAF square plane (Eq. \ref{2dHT}).

\ FIG. 6 \ Normalised specific heat $C_{p}/(N_{A}k_{B})$ of Na$_{2}$CuP$_{2}%
$O$_{7}$ is displayed as a function of temperature $T$. \ The inset has
$d[C_{p}/(N_{A}k_{B})]/dT$ as a function of $T$ showing an anomaly around $10$ K.

\ FIG. 7 \ The $^{31}$P nuclear spin-lattice relaxation rate $1/T_{1}$ vs
temperature $T$ for Na$_{2}$CuP$_{2}$O$_{7}$. The open circles are our
experimental results and the solid line represents the simulated curve of Eq.
\ref{2Dt1}. In the inset, the magnetization recoveries are plotted as a
function of pulse separation $t$ and the solid line is an exponential fit to
Eq. \ref{T1}.

\ FIG. 8 \ The $^{31}$P spin-spin relaxation rate $1/T_{2}$ is plotted as a
function of temperature $T$. In the inset, spin-echo decays are plotted as a
function of $t$ at three different temperatures for Na$_{2}$CuP$_{2}$O$_{7}$.
The solid lines show the fitting to a exponential function (Eq. \ref{T2}).


\begin{thebibliography}{99}
\bibitem {footnote}This temperature range is not low enough for the "$T^{3}$-
approximation" of the lattice heat capacity to be valid. We tried to determine
the lattice contribution to the specific heat by fitting the high-temperature
data (say, between $40$ K and $70$ K, where the magnetic contribution may be
expected to be negligible) to the exact expression of the Debye model.
\ However, this fit (not shown) deviated significantly from the data when
extrapolated to higher and lower temperatures. \ Considering larger ranges of
temperature for the fit, required us to use a temperature dependent (we tried
a second order polynomial form with coefficients as fitting parameters) Debye
temperature. \ This too did not appear to be helpful in reliably extracting
the magnetic contribution at low temperatures (below 15 K).
\end{thebibliography}
\end{document}